\documentclass[12pt]{iopart}
\usepackage{graphicx}
\usepackage{hyperref}
\hypersetup{
     colorlinks = true,
     linkcolor = blue,
     anchorcolor = blue,
     citecolor = blue,
     filecolor = blue,
     urlcolor = blue
     }

\begin{document}

\title[The Aubry-Andr\'e model for undestanding localization phenomenon]{The Aubry-Andr\'e model as the hobbyhorse for understanding localization phenomenon}

\author{G.A. Dom\'{\i}nguez-Castro \& R. Paredes}

\address{Instituto de F\'{\i}sica, Universidad
Nacional Aut\'onoma de M\'exico, Apartado Postal 20-364, M\'exico D.
F. 01000, Mexico}
\ead{gustavodomin@estudiantes.fisica.unam.mx}
\vspace{10pt}
\begin{indented}
\item[]December 2018
\end{indented}

\begin{abstract}
We present a thorough pedagogical analysis of the single particle localization phenomenon in a quasiperiodic lattice in one dimension. Description of disorder in the lattice is represented by the Aubry-Andr\'e model. Characterization of localization is performed through the analysis of both, stationary and dynamical properties. The stationary properties investigated are the inverse participation ratio (IPR), the normalized participation ratio (NPR) and the energy spectrum as a function of the disorder strength. As expected, the distinctive Hofstadter pattern is found. Two dynamical quantities allow to discern the localization phenomenon, being the spreading of an initially localized state and the evolution of population imbalance in even and odd sites across the lattice.
\end{abstract}

\section{Introduction}
The understanding of electronic mobility in quasiperiodic or disordered media is one of the fundamental issues in the condensed matter domain. Disorder, together with the interparticle interactions intrinsically present in every macroscopic sample, are the responsible ones of the physical behavior and response properties of solids. A material becomes insulator as a result of either electron-electron or electron-ion interactions. While the study of electron-electron interactions demands the use of a many body theory and leads to identify a Mott insulating phase \cite{Mott}, electron-ion interactions are addressed within the single electron theory and allow to discern among several types of insulators. Among them, band insulators arising from periodicity in the lattice \cite{Mermin}, Peierls insulators associated with lattice distortions, and Anderson insulators resulting from lattice imperfections or impurities, also known as lattice disorder. The purpose of this manuscript is to show how this loss of long range order in the lattice leads to localize single electrons and thus cause the absence of its diffusion across the lattice. Although localization phenomenon has been extensively investigated, is up to now a central topic which still have a manyfold of open questions to be addressed. For instance, the interplay between disorder and dimensionality, interactions and inhomogeneity created by external fields among others.

Characterization of localization phenomenon requires as a first step the election of an effective model representing the disorder, and then the use of the standard quantum mechanics techniques to analyze their effects. There are two general schemes from which localization has been envisaged, the Anderson model \cite{Anderson}, in which disorder is represented by a random amplitude of the on-site energies, and the Aubry-Andr\'e model \cite{Aubry, Harper}, where disorder arises from the superposition of two lattice potentials with incommensurate wavelengths. These models capture the metal-insulator transition in disordered lattices and allow to characterize such transition by tracking different properties as we describe below. Here, we shall use the Aubry-Andr\'e model as the hobbyhorse for studying and characterizing the influence that disorder has in producing localized states. This model introduced in 1980, has shown to be very successful in describing such transition not only in the single electron case but when interparticle interactions are considered \cite{Fisher, Kisker, Balabanyan, Gimperlein, Buonsante}.
Localization in a lattice can be recognized through several signatures, either of stationary or dynamical character. What it is important to stress is that localization can result from both, destructive interference associated with the multiple scattering of the wave function traveling along the disordered medium and the spectral properties of the Schr\"odinger equation \cite{Sarma}.  Destruction of wave coherence or loss of mobility is quantified in terms of several properties that can be extracted from the wave function. The first distinctive signature of localization, identified in the seminal work of Anderson, was the localization length that measures the size of the exponentially localized single particle state as a function of disorder strength.  Here we concentrate on analyzing the properties enunciated in the following lines. First, we analyze the properties of the Aubry-Andr\'e model, and then we investigate the inverse participation ratio (IPR) and its opposite, the normalized participation ratio (NPR), that quantify the fraction of sites contributing to the state along the lattice. Next, we investigate the energy spectrum that also allows monitoring the transition to localization. In addition to these quantities characterizing localization of the stationary states, there are dynamical parameters that also allow to track the evolution of a given initial state in the presence of disorder. Among them, the spreading of the initial state and the imbalance between the density probability of even and odd sites in the lattice, as a function of time.

Current experiments with ultracold neutral atoms realized in the laboratory represent the ideal scenario where the spatial quasiperiodicity of the Aubry-Andr\'e model can be recreated. Optical lattice potentials produced by standing waves resulting from interfering laser fields emulate such a non crystalline environment seen by electrons moving across the ion cores. Nowadays, such large ensembles of fermionic or bosonic atoms loaded in optical lattices offer advantages with respect to experiments performed in solids since ultracold atoms can be prepared to analyze isolated effects present in solids, without the influence of further outcomes \cite{Bloch1}. As a matter of fact, 10 years ago the Aubry-Andr\'e model was experimentally set in a laboratory for the first time \cite{Roati1}.  
 
The aim of the present manuscript is to present a pedagogical description of the Aubry-Andr\'e model to understand and characterize the localization phenomenon. Advanced undergraduate and graduate students should be able to follow this article with no difficulty. We believe that this material should give the appropriate tools and techniques to face and approach forefront problems including the many-body localization phenomenon. The manuscript is organized in 6 sections. First in section 2 we derive the Aubry-Andr\'e model demonstrating how quasiperiodicity in the potential gives rise to a cosine function incommensurate with the underlying periodic tight-binding 1D lattice. Then, in section 3 the properties characterizing the Aubry-Andr\'e model are delineated. Sections 4 and 5 account for the time independent and time dependent analysis that characterizes the localization transition. Finally, in section 6 a summary of results is presented.

\section{Model}
\begin{figure}[h]
\centering
\includegraphics[width=0.4\textwidth]{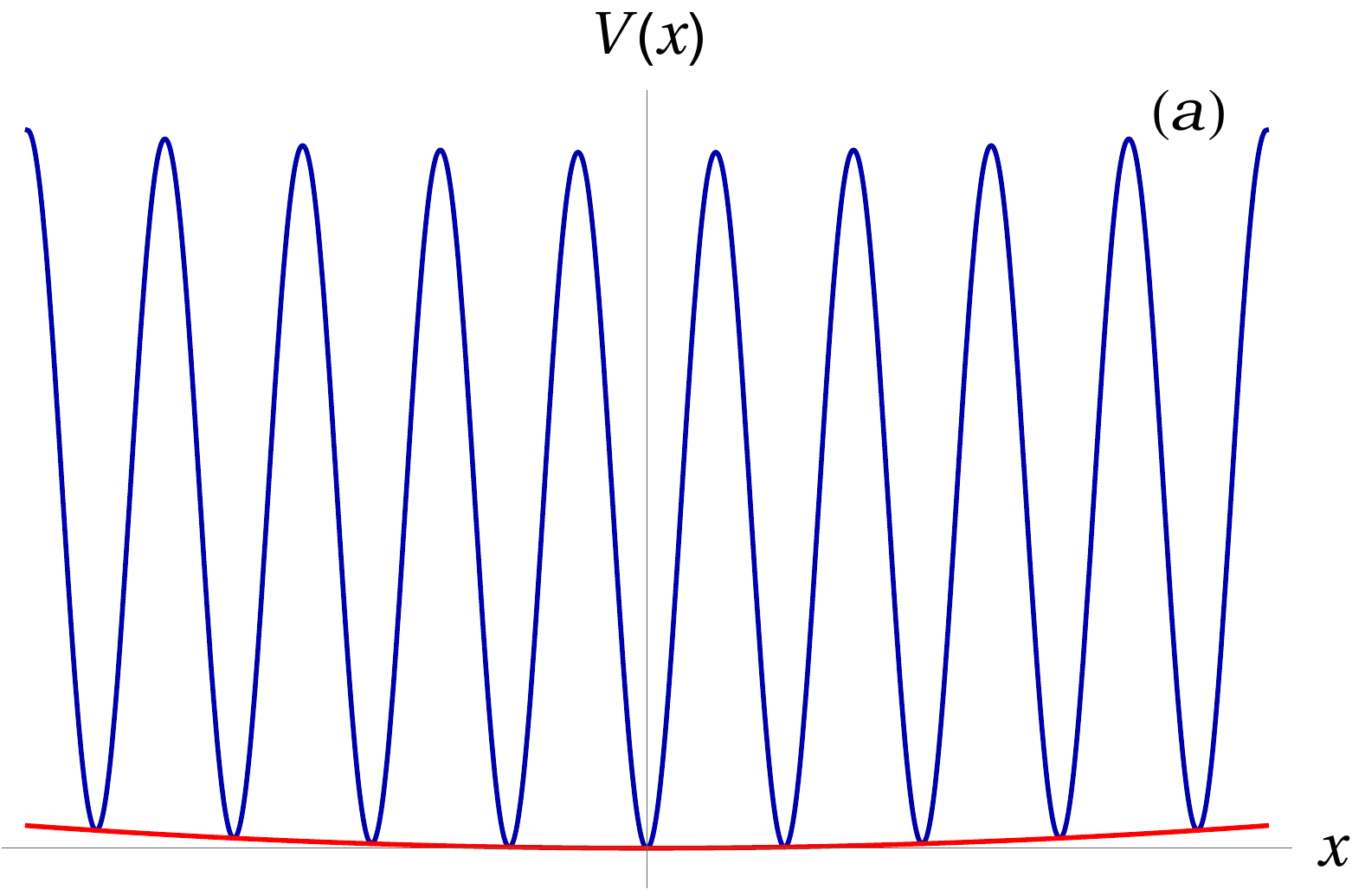}
\includegraphics[width=0.4\textwidth]{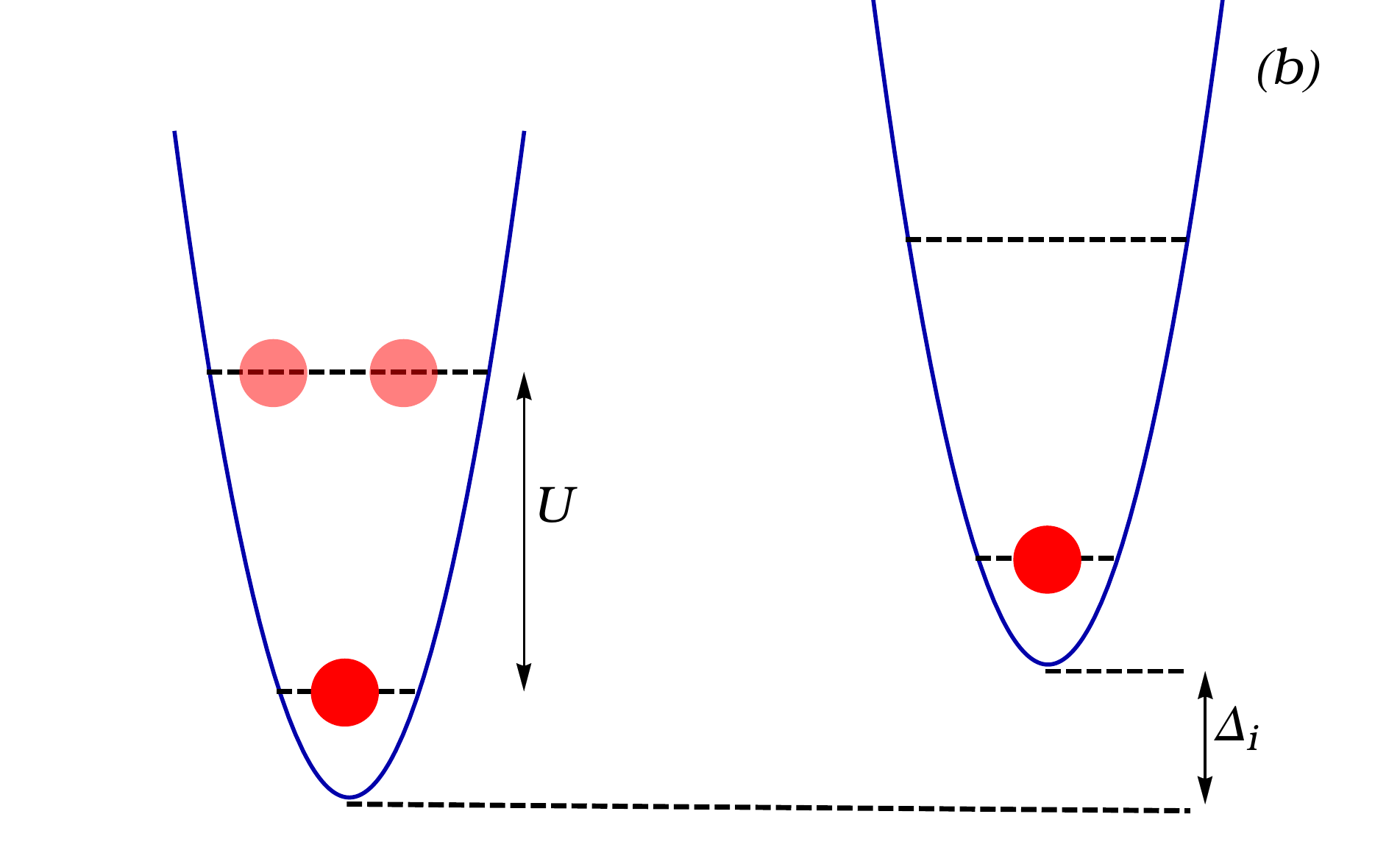}
\caption{(a) Sketch of the confining potential where the ultracold gas of Bose atoms move $V(x) = V_{T}(x)+V_{opt}(x)$. It results from adding both, the magnetic harmonic trap $V_{T}(x)$ and the optical lattice potential $V_{opt}(x)$. (b) A couple of nearest neighbor wells of (a) in the presence of disorder, the  site-to-site energy difference is $\Delta_{i}$.}
\label{Fig1}
\end{figure}

Our starting point is the 1D Hamiltonian operator for ultracold bosonic atoms with mass $m$ confined in an external potential  $V(x)$ and interacting via a contact potential written, as usual, in terms of the $s$- wave scattering length $a_s$,
\begin{eqnarray}
\fl
\hat{H} = \int dx \ \hat{\psi}^{\dagger}(x)\left( -\frac{\hbar^2}{2m}\nabla^{2} + V(x) \right)\hat{\psi}(x)\nonumber\\
+ \ \frac{1}{2} \ \frac{4\pi a_{s}\hbar^{2}}{m}\int dx \ \hat{\psi}^{\dagger}(x)\hat{\psi}^{\dagger}(x)\hat{\psi}(x)\hat{\psi}(x),
\label{eq1}
\end{eqnarray}
$\hat{\psi}^{\dagger}(x)$ and $\hat{\psi}(x)$ are the bosonic creation and annihilation field operators satisfying the standard commutations rules for bosons, the external potential $V(x)$ is given by the superposition  $V(x) = V_{T}(x)+V_{opt}(x)$, in which $V_{T}(x)$ is a slowly varying magnetic harmonic trap and,  $V_{opt}(x)$ an optical lattice potential. \Fref{Fig1}(a) shows a sketch of the resulting potential $V(x)$. In presence of disorder, $V_{opt}(x)$ consists of two optical lattices \cite{Modugno}, the main lattice $V_{1}(x)=s_{1}E_{R_{1}}\sin^{2}(k_{1}x)$ which is used to create a tight-binding environment for the atoms and a secondary one $V_{2}(x)=s_{2}E_{R_{2}}\sin^{2}(k_{2}x)$ which introduces an optical disorder \cite{Fallani}. Superimposing both lattices gives rise to the following bichromatic potential:
\begin{eqnarray}
V_{opt}(x)&=V_{1}(x)+V_{2}(x)= s_{1}E_{R_{1}}\sin^{2}(k_{1}x) + s_{2}E_{R_{2}}\sin^{2}(k_{2}x+\varphi)\nonumber\\
&= s_{1}E_{R_{1}}\sin^{2}(k_{1}x)  + s_{2}E_{R_{1}}\beta^2\sin^{2}(\beta k_{1}x+\varphi)
\label{eq2}
\end{eqnarray}
where $k_{i}=2\pi/\lambda_{i}$  (i=1,2) are the wave vectors, with $\lambda_{i}$ the wavelength of the lasers fields, $s_{i}$ are the heights of the lattices in units of the recoil energy $E_{R_{i}}=h^{2}/(2m\lambda_{i}^{2})$, $\varphi$ is an arbitrary phase and $\beta=\lambda_{1}/\lambda_{2}$ the ratio between the wavelengths. When $s_{2}<<s_{1}$ and $\beta$ is an incommensurate number, the secondary lattice does not change considerably the positions of the potential minima generated by the main lattice \cite{Guarrera}. Instead, as shown in \fref{Fig1}(b) it has the effect of shifting the local site energy by an amount $\Delta_{i}$ only.

\begin{figure}[h]
\centering
\includegraphics[width=0.4\textwidth]{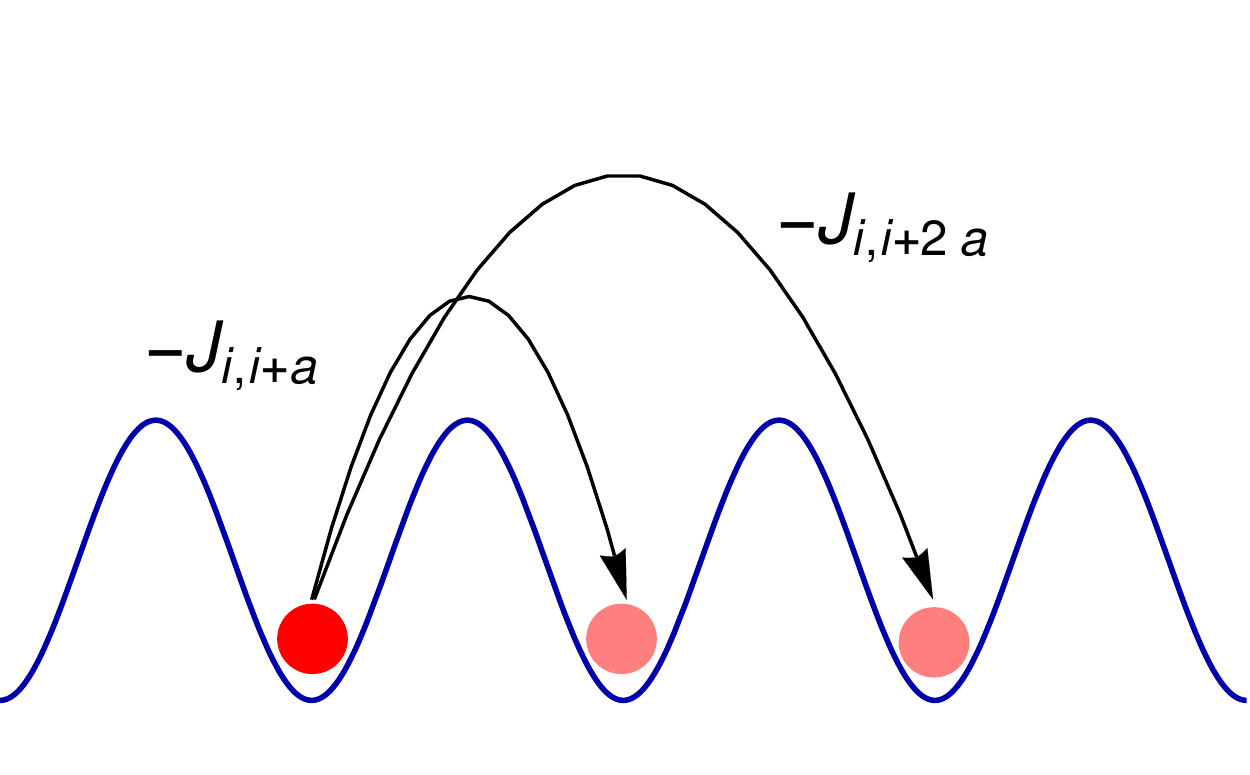}
\caption{Schematic representation of the hopping of a particle  in a lattice.}
\label{Fig2}
\end{figure}

For single atoms and no net disorder $s_{2} = 0$, the eigenstates of equation (\ref{eq1}) are Bloch wave functions \cite{Nolting}. As it is well known, an appropriate linear combination of Bloch states yields a Wannier wave function $w_{\nu}(x-x_{i})$, characterized by large probability amplitude around lattice site $i$, that is, a localized wave function at each site $i$. Since the atoms under study are at ultracold temperatures, it is well justified the assumption that the energies involved in the system are smaller compared to the energy required to allow second and higher band populations. This consideration allows us to drop the band index $\nu$ in the Wannier functions and contemplate first band populations only. Having this assumption in mind, it is convenient to expand the field operators $\hat{\psi}(x)$ and $\hat{\psi}^{\dagger}(x)$ in the Wannier basis:
\begin{eqnarray}
\eqalign{
\hat{\psi}(x) = \sum_{i}\hat{b}_{i}w(x-x_{i}),\cr
\hat{\psi}^{\dagger}(x) = \sum_{i}\hat{b}^{\dagger}_{i}w^{*}(x-x_{i}).
}
\label{eq3}
\end{eqnarray}
Being $\hat{b}_{i}$ and $\hat{b}^{\dagger}_{i}$ the annihilation and creation operators for a particle in a Wannier state at the lattice site $i$ respectively. It is worthwhile to stress that the sums in equation (\ref{eq3}) run over all lattice sites. As mentioned above, for weak disorder $s_{2}<<s_{1}$ the minima of the main lattice are not remarkably affected and we can safely substitute the latter expansion of the field operators in equation (\ref{eq1}). After some straightforward algebra, one can obtain the following expression for the Hamiltonian (\ref{eq1}):
\begin{eqnarray}
\fl
\hat{H} = -\sum_{i,j} J_{ij}\hat{b}_{i}^{\dagger}\hat{b}_{j} + 
\sum_{ij}\epsilon_{ij}\hat{b}_{i}^{\dagger}\hat{b}_{j}+\sum_{ij}\Delta_{ij}\hat{b}_{i}^{\dagger}\hat{b}_{j}+
\sum_{i,j,l,v} U_{i,j,l,v} \  \hat{b}_{i}^{\dagger}\hat{b}_{j}^{\dagger}\hat{b}_{l}\hat{b}_{v}, 
\label{eq4}
\end{eqnarray}
where we have defined the following constants:
\begin{equation}
\eqalign{J_{ij} = -\int dx \  w^{*}(x-x_{i})\left(-\frac{\hbar^2}{2m}\nabla^2 + s_{1}E_{R_{1}}\sin^2(k_{1}x)\right)w(x-x_{j})\cr
\epsilon_{ij} = \int dx \  w^{*}(x-x_{i})V_{T}(x)w(x-x_{j})\cr
\Delta_{ij} = s_{2}E_{R_{1}}\beta^{2}\int dx \  w^{*}(x-x_{i})\sin^{2}(\beta k_{1}x+\varphi)w(x-x_{j})\cr
U_{ijlv} = \frac{4\pi a_{s}\hbar^2}{2m} \int dx \ w^{*}(x-x_{i})w^{*}(x-x_{j})w(x-x_{l})w(x-x_{v}).} 
\label{eq5}
\end{equation}
The first term in equation (\ref{eq4}) describes the energy cost for an atom to hop from site $i$ to site $j$ with $i\neq j$, see \fref{Fig2}.
Note that the hopping probability $J_{ij}$ is proportional to the overlap between the Wannier functions centered at different lattice sites. Within the so-called tight-binding approximation, this overlap is essential only for nearest neighbors \cite{Nolting}, thus we can neglect the tunneling between next nearest neighbors and beyond.
Also, since the main lattice potential is invariant under translations by one lattice period, $a=\lambda/2$, the hopping parameter $J_{ij}$ becomes a constant $J$ independent of the lattice site. For the case $i=j$ the $J_{ii}$ term represents an on-site energy shift which is equal for all sites and thus can be dropped. Also, the second term $\epsilon_{ij}$ represents an on-site shift of the energy. However, this time we have to assume that the frequency $\omega_{T}$ of the harmonic trap satisfies \cite{Jaksch} $\sqrt{\hbar/(m\omega_{T}})<<\lambda_{1}/2$ which allow us to consider the variation of the energy from site to site as equal, thus to a first approximation, we can neglect the contribution of the harmonic trap, see \fref{Fig1}(a). The third term in equation (\ref{eq4}) is the responsible for the optical disorder in the lattice. In order to deal with it, we first use the trigonometric relation $\sin^{2}(\beta k_{1}x + \varphi) = (1-\cos(2\beta k_{1}x+\varphi'))/2$ with $\varphi'=2\varphi$. Inserting this relation into the third equality of equation (\ref{eq5}) and dropping the constant term, we obtain:
\begin{eqnarray}
\Delta_{ij} = -\frac{s_{2}E_{R_{1}}\beta^{2}}{2}\int  dx \  w^{*}(x-x_{i})\cos(2\beta k_{1}x+\varphi')w(x-x_{j}).
\label{eq6}
\end{eqnarray}
Again, for deep enough lattices, the leading contribution of equation (\ref{eq6}) is the $i=j$ term, which corresponds to an on-site energy shift variation. Further, we can make the change of variable $y=x-x_{i}$, leading to:
\begin{eqnarray}
\Delta_{ii} = -\frac{s_{2}E_{R_{1}}\beta^{2}}{2}\cos(2\pi\beta i + \varphi')\int dy \ \cos(2\beta k_{1}y)|w(y)|^{2}.
\label{eq7}
\end{eqnarray} 
Where in the last equation we have identified $x_{i}\rightarrow i$, used the subsequent trigonometric identity: 
\begin{eqnarray}
\eqalign{
\cos(2\beta k_{1}y+2\beta k_{1}x_{i}+\varphi') =& \cos(2\pi\beta i + \varphi')\cos(2\beta k_{1}y)-\cr &\sin(2\pi\beta i + \varphi')\sin(2\beta k_{1}y),}
\label{eq8}
\end{eqnarray}
and symmetric properties to drop the sine integral. Following the above steps we finally get the usual disorder term \cite{Modugno}:
\begin{eqnarray}
\Delta_{ij} = \Delta\cos(2\pi\beta i + \phi)\delta_{ij},
\label{eq9}
\end{eqnarray}
where $\phi = \varphi' + \pi$ and $\Delta$ is defined as the following constant parameter:
\begin{eqnarray}
\Delta = \frac{s_{2}E_{R_{1}}\beta^{2}}{2}\int dy \cos(2\beta k_{1}y)|w(y)|^{2}.
\label{eq9a}
\end{eqnarray}
The term associated with the interaction energy can also be simplified by taking into account the tight binding approximation. The dominant term of the overlap of four Wannier functions is due to the term $i=j=l=v$, which corresponds to an onsite interaction, where the atoms only ``see each other" whenever they are in the same lattice site. 
\begin{eqnarray}
U = U_{i,i,i,i} = \frac{4\pi a_{s}\hbar^2}{2m}\int |w(x)|^4 dx.
\label{eq10}
\end{eqnarray} 
Summarizing all the above approximations, we end with the following interacting Hamiltonian:
\begin{eqnarray}
\hat{H} = -J\sum_{\langle i, j\rangle} \hat{b}_{i}^{\dagger}\hat{b}_{j} + \Delta\sum_{i}\cos(2\pi\beta i + \phi)\hat{n}_{i}+U\sum_{i}\hat{n}_{i}(\hat{n}_{i}-1),
\label{eq11}
\end{eqnarray}
where the notation $\langle i, j\rangle$ indicates that the sum runs over nearest neighbors only and $\hat{n}_{i} = \hat{b}_{i}^{\dagger}\hat{b}_{i}$ is the number operator at site $i$. One of the most outstanding advantages of the experiments with ultracold atomic gases is the possibility of tuning the strength of the pairwise interactions between atoms, via an external magnetic field. This procedure called Feshbach resonance \cite{Bloch1} allows the experimentalist to prepare a gas of atoms with a zero scattering length $a_s$, and consequently $U = 0$. Such a non-interacting system constitutes an experimental realization of the non-interacting Harper \cite{Harper} or Aubry-Andr\'e \cite{Aubry} model:
\begin{eqnarray}
\hat{H} = -J\sum_{\langle i, j\rangle} \hat{b}_{i}^{\dagger}\hat{b}_{j} + \Delta\sum_{i}\cos(2\pi\beta i + \phi)\hat{n}_{i}.
\label{eq12}
\end{eqnarray}
This model and also the interacting version has been realized in experiments with ultracold atoms in bichromatic optical lattices potentials \cite{Roati1,Bloch3}. Written in the Dirac notation, the above Hamiltonian takes the form:
\begin{eqnarray}
\fl
\hat{H} = -J\sum_{j} (|w_{j}\rangle\langle w_{j+1}| + |w_{j+1}\rangle\langle w_{j}|) + \Delta\sum_{j}\cos(2\pi\beta j + \phi)|w_{j}\rangle\langle w_{j}|.
\label{eq13}
\end{eqnarray}
It is important to emphasize that recent investigations \cite{Sarma,Bloch2} pointed out that the Hamiltonian in equation (\ref{eq13}) is strictly valid in the extreme tight-binding limit of a very deep main lattice potential.

\section{Properties of the Aubry-Andr\'e model}
Having set the Aubry-Andr\'e Hamiltonian it is worthwhile to expose some basic properties of this model. Beginning with its duality in space and momentum representations, one can transform the Hamiltonian (\ref{eq13}) written in the Wannier representation to one in the momentum representation, via the following transformation:
\begin{eqnarray}
|k_{s}\rangle = \sum_{j} e^{2\pi i\beta k_{s}j}|w_{j}\rangle
\label{eq14} 
\end{eqnarray}
After substitution and straightforward algebra, we find the dual Hamiltonian:
\begin{eqnarray}
H = -\frac{\Delta}{2J}J\sum_{s} (| k_{s}\rangle\langle k_{s+1}| + |k_{s+1}\rangle\langle k_{s}|) + \frac{2J}{\Delta}\Delta\sum_{s}\cos(2\pi\beta s)|k_{s}\rangle\langle k_{s}|,
\label{eq15}
\end{eqnarray}
which has the same structure as that of (\ref{eq13}) except that the tunneling rate has changed from $J\rightarrow \Delta/2$ and the disorder strength from $\Delta\rightarrow 2J$. Also, for simplicity we set $\phi = 0$. Now, according to the Heisenberg uncertainty principle the Hamiltonian (\ref{eq15}) has localized states where the first Hamiltonian (\ref{eq13}) has extended states, and viceversa. Obviously, the transition from extended to localized states or from localized to extended  states must take place at the same set of parameters for both Hamiltonians. This reasoning leads us to impose a condition over the two parameters characterizing both Hamiltonians, the condition at the transition point is $\Delta/J=2$. Furthermore, the localization transition is sensitive to the value of the parameter $\beta$ \cite{Modugno}. For instance, an integer value $\beta$ would not display any localization transition. In order to observe such a transition, $\beta$ must have some degree of incommensurability. One way to achieve this requirement is to choose $\beta$ as the ratio of two adjacent Fibonacci numbers $F_{n-1}/F_{n}$ \cite{Fibonacci}, with $F_{n-1}$ and $F_{n}$ two consecutive elements of the Fibonacci sequence. Such a procedure approaches the inverse Golden ratio $\varphi = (\sqrt{5}-1)/2$ for large enough Fibonacci numbers. Also, one can express $\beta$ as the ratio of two relative prime numbers $\beta=P/Q$ with $P$ and $Q$ larger than the system size of simulation \cite{Modugno2}. However, in experimental realizations, $\beta$ is restricted to the available laser wavelengths. For instance, recent experiments \cite{Bloch2} were performed with $\beta=532/738$. Summarizing, we can say that we must ensure that $\beta$ is such that the system remains aperiodic within the size of interest \cite{Sarma}. If these requirements are accomplished one would observe that for $\Delta/J = 2$, the wave function develops peaks around certain lattice sites, such peaks become a single peak as the disorder amplitude $\Delta$ is increased. As shown by Thouless \cite{Thouless}, there exists a link between the density of states $\rho$ and the inverse of the localization length $l$ of the wave function. In general, the localization length depends on the energy of the localized wave function. One can find such dependence by solving the following integral:
\begin{eqnarray}
l(E)^{-1} = \int_{-\infty}^{\infty} \ln |E-\epsilon |\ d\rho(\epsilon).
\label{eq16}
\end{eqnarray} 
For the Aubry-Andr\'e model considered here, the latter equation gives an energy independent localization length of $l=d/\ln(\Delta/2J)$ \cite{Modugno2} which implies that all eigenstates would display the same localization length.\\
Another remarkable property of the Aubry-Andr\'e model is related to the type of disorder generated, which is called correlated disorder. This is because, the correlation function of the disorder potential $G(y)=\langle V(x)V(x+y)\rangle$ takes the value:
\begin{eqnarray}
G(y) = \frac{\Delta^{2}}{2\pi}\cos(2\pi\beta y).
\label{eq17}
\end{eqnarray}
This disorder correlation is in contrast with the uncorrelated Anderson disorder \cite{Anderson} where any random disorder localizes the system. 

\section{Time independent results}
\begin{figure}[h]
\centering
\includegraphics[width=0.45\textwidth]{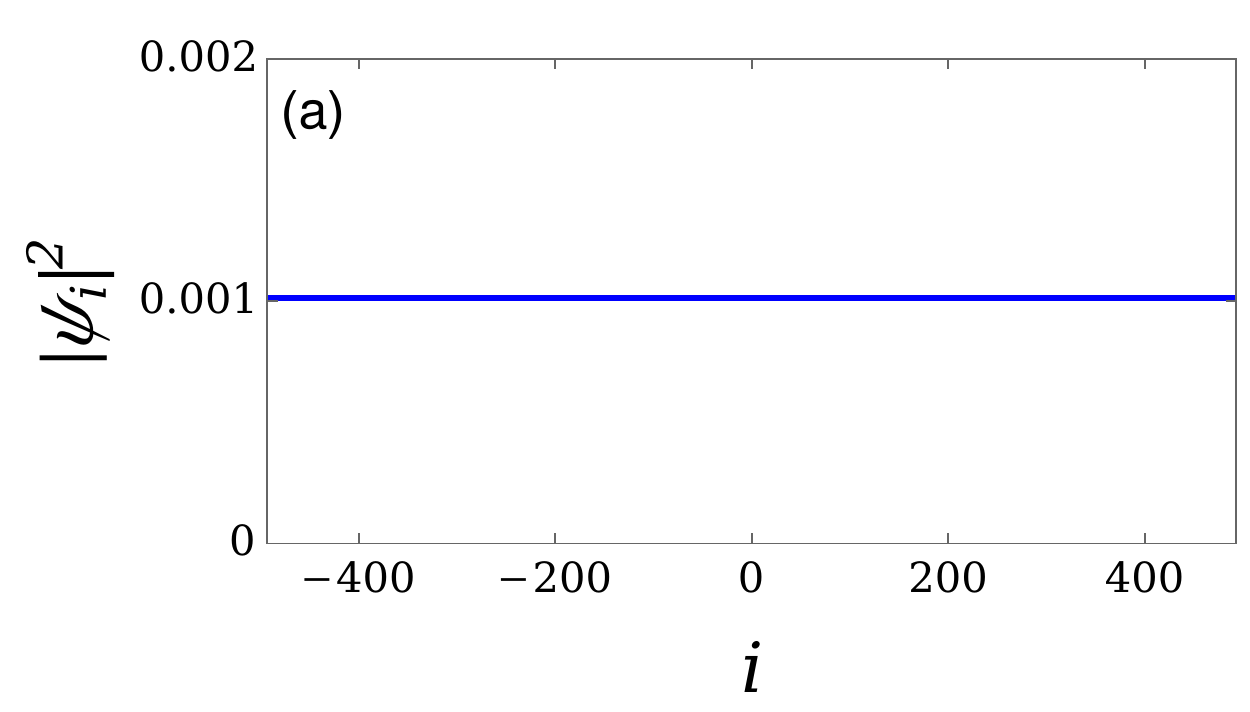}
\includegraphics[width=0.45\textwidth]{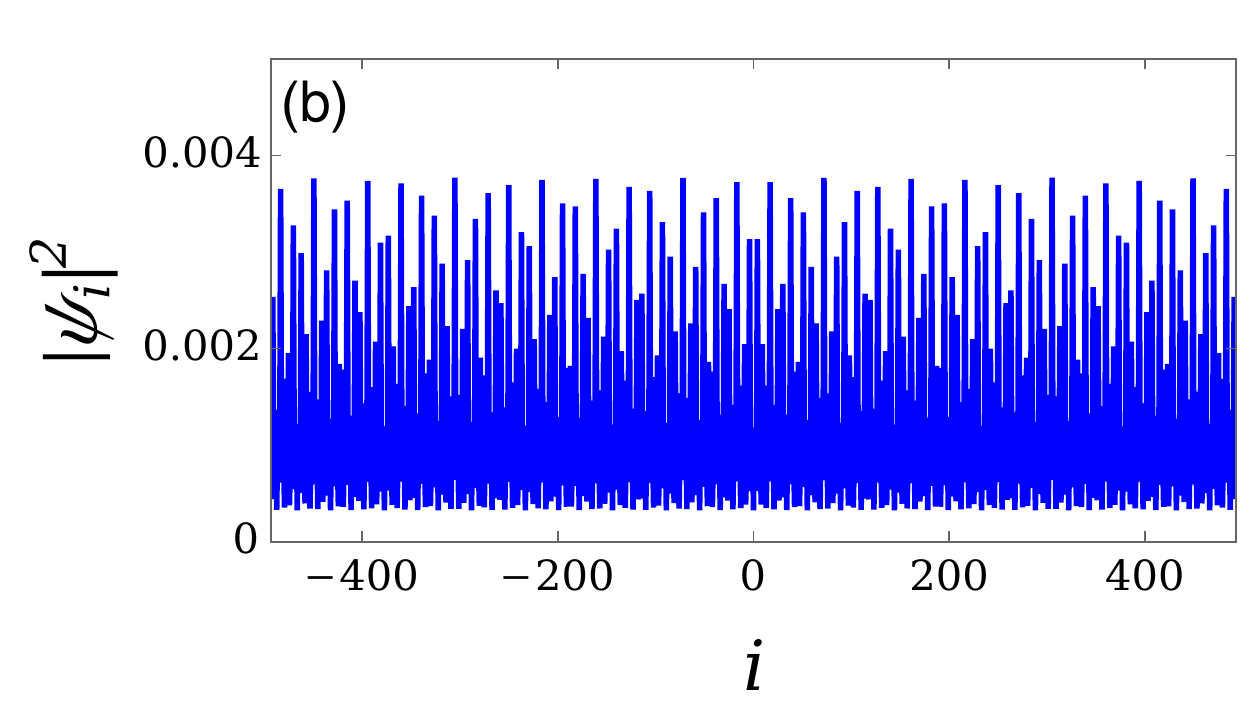}
\caption{(a) Square of the ground state wave function $\psi_{i}$ versus the lattice site index $i$ for $\Delta/J = 0$ and $\phi=\pi/5$. (b) Square of the ground state wave function $\psi_{i}$ versus the lattice site index $i$ for $\Delta/J = 1.5$ and $\phi=\pi/5$}
\label{Fig3}
\end{figure}

In this section, we describe some of the most distinctive single particle time independent results of the Aubry-Andr\'e model. In the deep tight-binding approximation, the Wannier functions are highly localized and thus can be represented by the site basis \cite{Nolting}, that is, $|w_{i}\rangle\rightarrow|x_{i}\rangle$. In this scenario, the equation (\ref{eq13}) is reduced to the Schr\"odinger equation $\hat{H}|\psi\rangle=E|\psi\rangle$ with $|\psi\rangle$ being an eigenstate of the one particle Hamiltonian $\hat{H}$:
\begin{eqnarray}
\hat{H} = -J\sum_{i}(|x_{i}\rangle\langle x_{i+1}|+|x_{i+1}\rangle\langle x_{i}|)+\Delta\sum_{i}\cos(2\pi\beta i+\phi)|x_{i}\rangle\langle x_{i}|
\label{eq18}
\end{eqnarray}
By expanding the wave function $|\psi\rangle$ in terms of the site basis $|\psi\rangle = \sum_{i}\psi_{i}|x_{i}\rangle$ and calculating the product $\langle x_{j}|\hat{H}|\psi\rangle$, we obtain the following difference equation for the coefficients $\psi_{i}$:
\begin{eqnarray}
-(\psi_{j+1}+\psi_{j-1}) + \frac{\Delta}{J}\cos(2\pi\beta j+\phi)\psi_{j} = \frac{E}{J}\psi_{j},
\label{eq19}
\end{eqnarray}
where we have divided by $J$ in order to have dimensionless equations and at the same time to set the energy scale. It is easy to see that equation (\ref{eq19}) can be rewritten in a matrix form $\mathbf{H}\vec{\psi}=\frac{E}{J}\vec{\psi}$ being $\vec{\psi} = (\psi_{1},\psi_{2},\cdots,\psi_{\Omega})^{T}$ the state vector and $\mathbf{H}$ the Hamiltonian matrix:

\begin{eqnarray}
\fl
\mathbf{H} =
\left(\begin{array}{ccccc}
      \Delta/J \cos(2\pi\beta+\varphi)&-1&\cdots &-1\\
      -1&\Delta/J \cos(4\pi\beta+\varphi)&\cdots &0\\
      \vdots & \ddots & \ddots & -1\\
      -1 &0&-1&\Delta/J \cos(2\Omega\pi\beta+\varphi)
\end{array}\right),
\label{eq20}
\end{eqnarray}

\noindent where $\Omega$ is the number of sites in the lattice. In the following calculations we consider $\Omega=987$ sites, although similar results are found for larger lattice sizes. At this point, it is instructive to state two aspects that were considered in our numerical calculations. The first one is related to the boundary conditions of the problem. As can be seen from the Hamiltonian matrix (\ref{eq20}), we take periodic boundary conditions, which means that the 1D lattice closes itself. Also, for the subsequent calculations we consider $\beta = 610/987 \approx 0.618034$.\\
Now, we are left to find the eigenvalues and eigenvectors of the matrix $\mathbf{H}$. In the case of vanishing disorder $\Delta/J = 0$, the equation (\ref{eq19}) is easily solved with the ansatz $\psi_{j} = e^{ikaj}$, which displays the energy spectrum of a free particle in a 1D lattice $E_{k} = -2J\cos(ka)$ \cite{Mermin}. In \fref{Fig3}(a) we plot the ground state density $|\psi_{i}|^{2}$ as a function of the lattice site $i$ for zero disorder $\Delta/J = 0$ and nonzero disorder $\Delta/J = 1.5$ in \fref{Fig3}(b). 

As one would expect, the ground state profile in the absence of disorder is a normalized constant at each site. This means that the particle is completely delocalized in the lattice. For the case $\Delta/J = 1.5$, the density profile of the ground state displays multiple peaks which indicates the presence of different potential depths across the sites. However, the wave function is still extended over all the lattice.

\begin{figure}[h]
\centering
\includegraphics[width=0.45\textwidth]{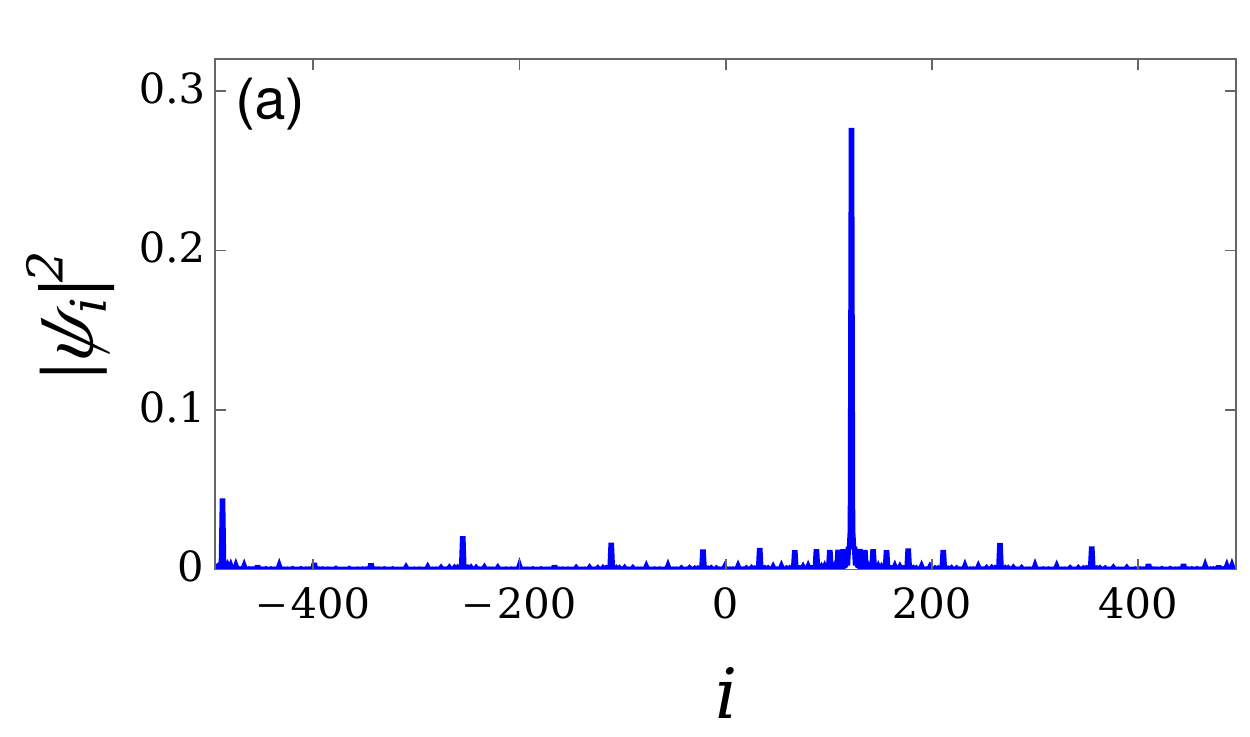}
\includegraphics[width=0.45\textwidth]{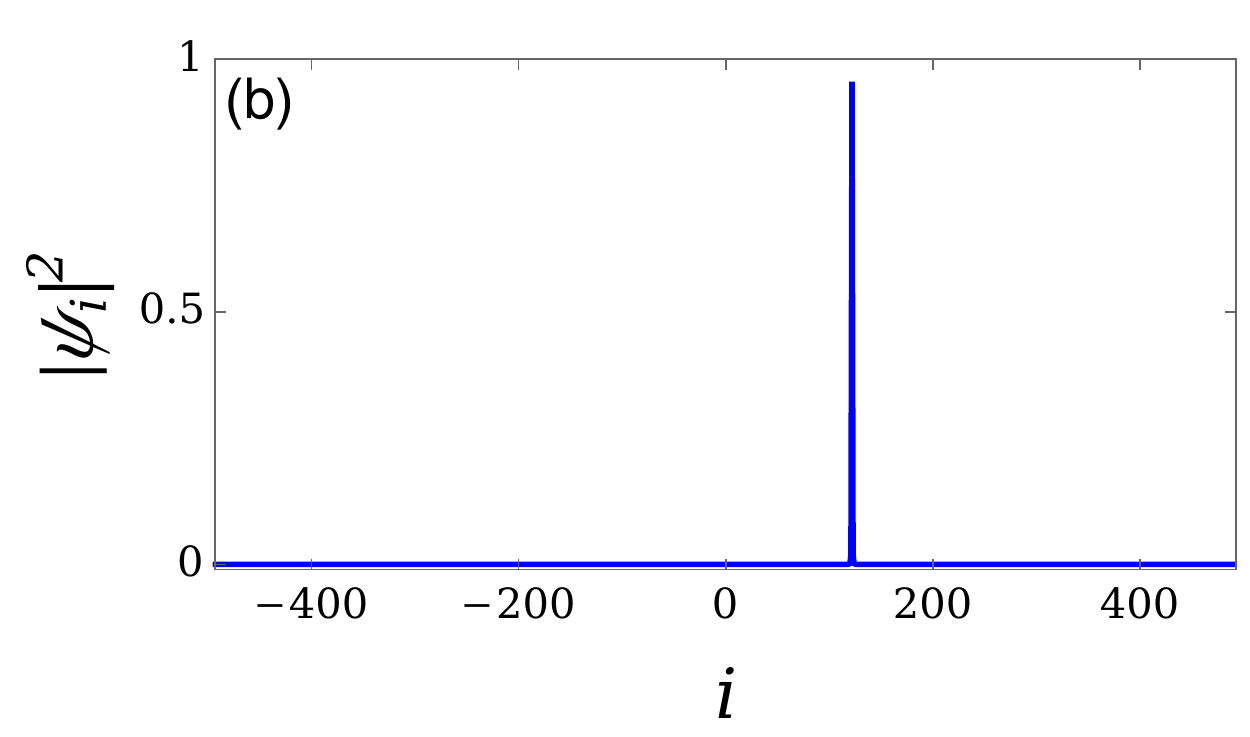}
\caption{(a) Density profile versus lattice index $i$ for $\Delta/J = 2$ and $\phi=\pi/5$. (b) Density profile as a function of the lattice index $i$ for $\Delta/J = 4$ and $\phi = \pi/5$.}
\label{Fig4}
\end{figure}

With aim of sketching the ground state density for two different values of the disorder amplitude satisfying $\Delta/J\geq 2$, in \fref{Fig4} we exhibit two cases, the left one associated to $\Delta/J = 2$ and the right one corresponding to $\Delta/J = 4$. As can be seen from these density profiles,  $\Delta/J = 2$ exhibits the transition from an extended to a localized state, while the case $\Delta/J = 4$ shows that when the disorder amplitude is increased, the localization becomes sharply.

An important quantity, that arises in describing the localization transition, and is widely used in the literature, is the inverse participation ratio (IPR). For a normalized state $|\psi\rangle=\sum_{i}\psi_{i}|x_{i}\rangle$, it is defined as:
\begin{eqnarray}
\mathrm {IPR}(|\psi\rangle) = \sum_{i}|\psi_{i}|^4.
\label{eq21}
\end{eqnarray}
The IPR gives us the inverse of the number of sites occupied by the wave function. For instance, it approaches zero as $1/\Omega$, for an extended wave function, while it goes to $1$ for a localized state on a single lattice site.
 
\begin{figure}[h]
\centering
\includegraphics[width=0.54\textwidth]{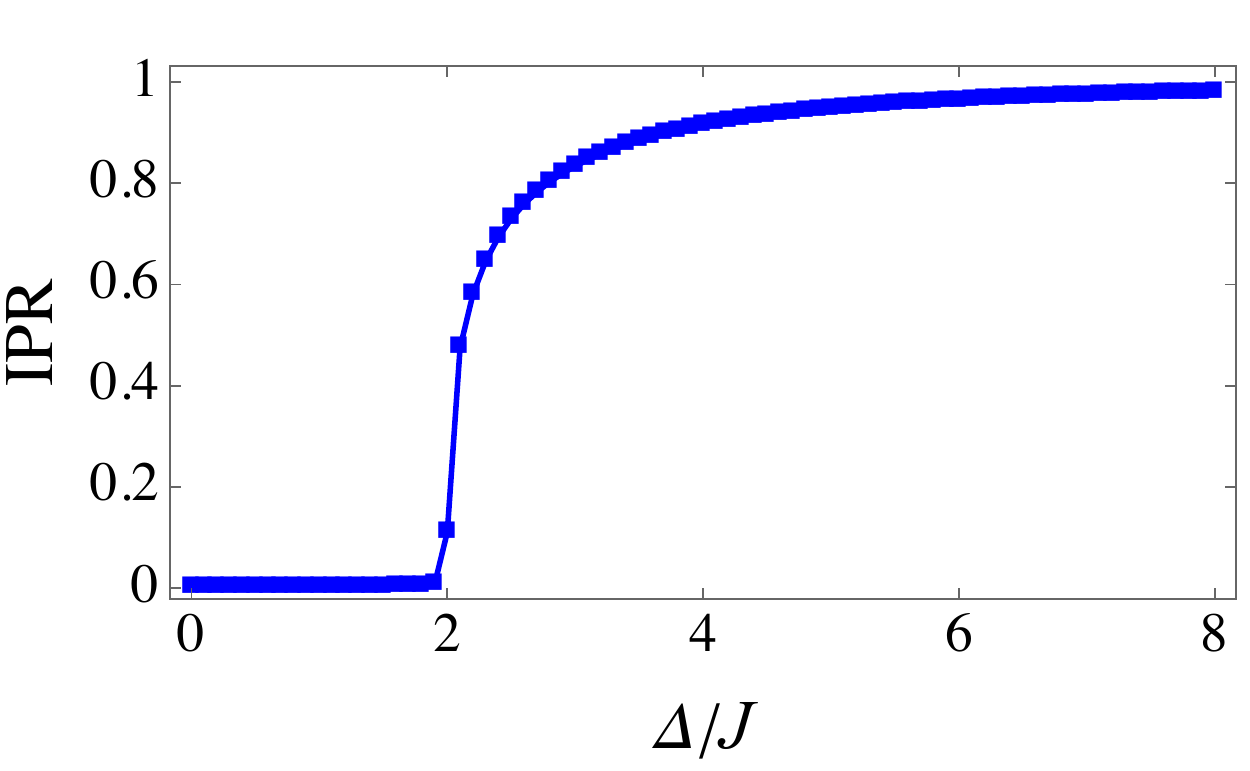}
\caption{The ground state inverse participation ratio as a function of the disorder strength $\Delta/J$. We average over $10$ realizations of the phase $\phi$.}
\label{Fig5}
\end{figure} 
 
The great advantage of using the IPR parameter is that instead of looking at the wave function in each realization, we just have to check a single parameter to confirm the nature of the wave function. In \fref{Fig5} we illustrate the IPR associated with the ground state as a function of the disorder strength $\Delta/J$. Each point in this figure corresponds to an average over 10 realizations of the phase $\phi$ which is set randomly in the interval $\phi\in [0,2\pi)$. As it is shown, the IPR parameter becomes different from zero for $\Delta/J = 2$ and it approaches the unity as the disorder increases. This peculiar behavior makes the IPR a good parameter for studying the localization transition. Even though the above definition of the IPR is related to the localization in real space, one could extend the idea to momentum space or more exotic basis as the Floquet basis in periodic driven optical lattices \cite{Qin,Pedro}. 
 
\begin{figure}[h]
\centering
\includegraphics[width=0.54\textwidth]{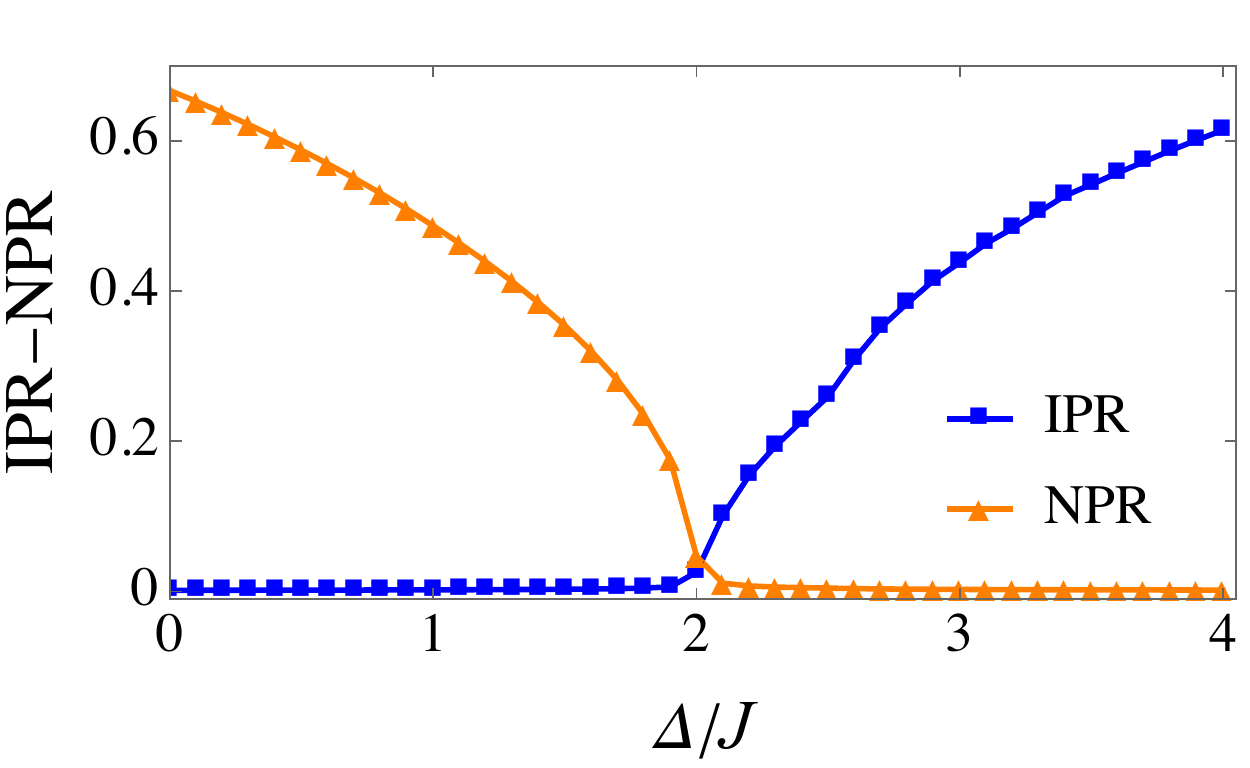}
\caption{Inverse participation ratio (IPR) and normalized participation ratio (NPR) of the full spectrum as a function of the disorder strength $\Delta/J$. Each point corresponds to the average of the full spectrum over $10$ realizations of the phase $\phi$.}
\label{Fig6}
\end{figure}

Another interesting quantity is the normalized participation ratio (NPR), which plays the opposite role of the IPR parameter \cite{Sarma}. That is, the NPR parameter remains finite for spatially extended states, while approaching zero for a localized one. For a given normalized state $|\psi\rangle=\sum_{i}\psi_{i}|x_{i}\rangle$, the NPR parameter is defined as follows:
\begin{eqnarray}
\mathrm{NPR}(|\psi\rangle) = \frac{1}{\Omega} \ \frac{1}{\sum_{i} |\psi_{i}|^{4}}.
\label{eq22}
\end{eqnarray}
In order to illustrate the utility of the NPR parameter, first, one should notice that the above definitions for the IPR and NPR parameters are related to a single eigenstate $|\psi\rangle$. However, one can calculate these two quantities for the full eigenstate spectrum, and display the average of the IPR and NPR parameters. In \fref{Fig6} we plot such averages as a function of the disorder strength $\Delta/J$.
As it is shown in this figure, for $\Delta/J<2$ we obtain, as expected, IPR $= 0$ and NPR $\neq 0$, while for $\Delta/J>2$, IPR $\neq 0$ and NPR$ = 0$. This indicate that the spectrum is either, completely delocalized ($\Delta/J<2$) or completely localized ($\Delta/J>2$), but not a mixture of both, localized and extended states.

\begin{figure}[h]
\centering
\includegraphics[width=0.38\textwidth]{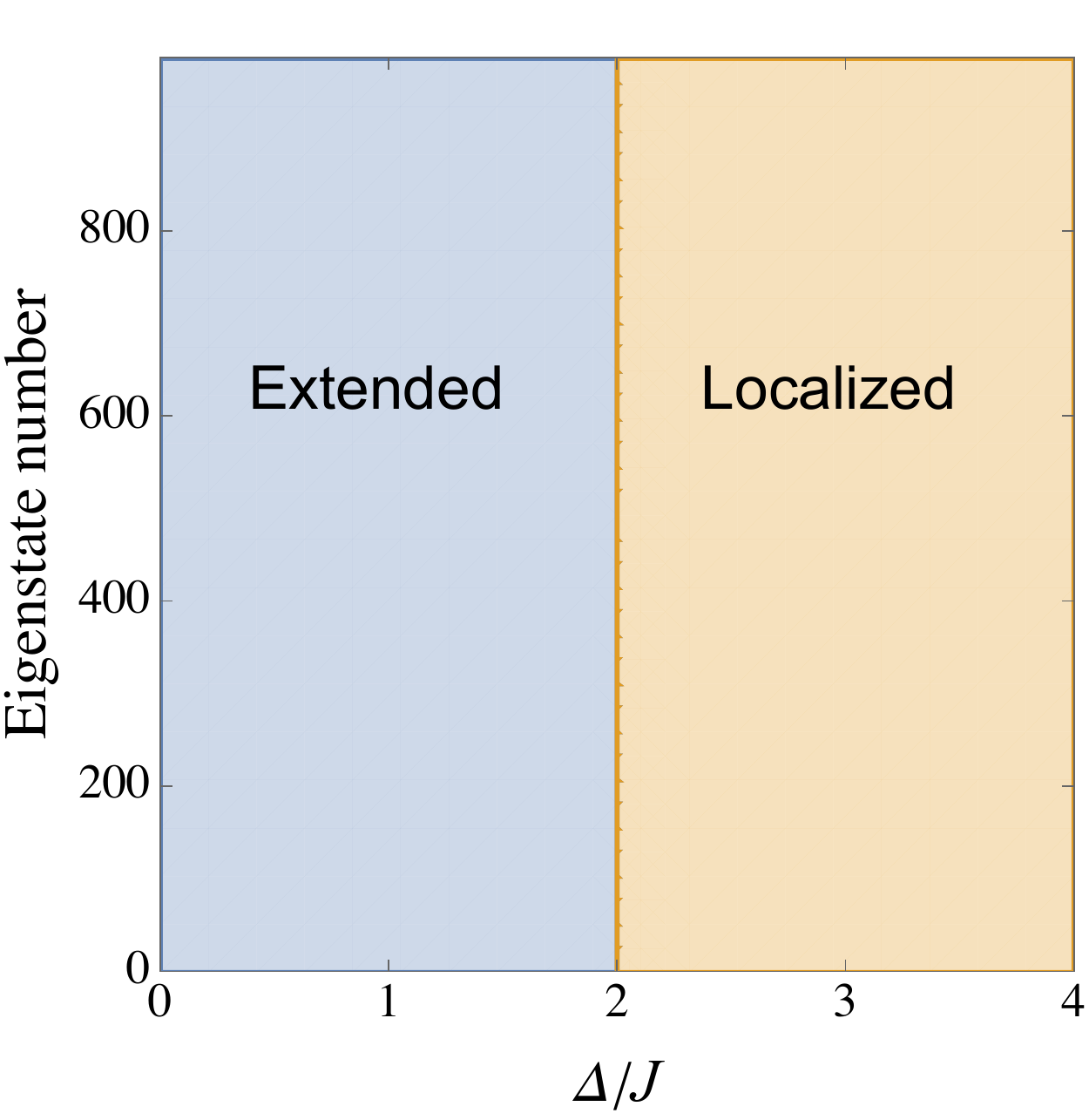}
\caption{Diagram of the Aubry-Andr\'e spectrum as a function of the disorder strenght $\Delta/J$.}
\label{Fig7}
\end{figure}

We should emphasize that this peculiar behavior of IPR and NPR curves, is a direct consequence of the tight-binding approximation used in deriving the Aubry-Andr\'e model \cite{Sarma}. Nevertheless, when hopping to next nearest neighbors is considered, a noticeable overlap between both curves emerges \cite{Sarma,Sarma2,Das,Yan}. In such scenario extended and localized states take place in the same spectrum, the value of the energy that separates localized and delocalized eigenstates is called mobility edge energy. At the transition point $\Delta/J=2$ all the eigenstates exhibit a multifractal structure \cite{Wilkinson}, a subject which is out of the scope of this work. Summarizing, we can organize the latter result in the diagram shown in \fref{Fig7} which displays an absence of mixture between localized and extended eigenstates in the full spectrum.

\begin{figure}[h]
\centering
\includegraphics[width=0.4\textwidth]{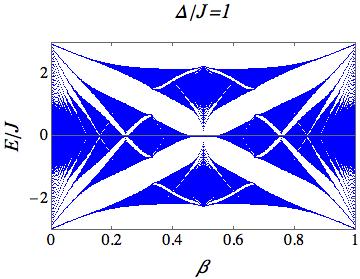}
\includegraphics[width=0.4\textwidth]{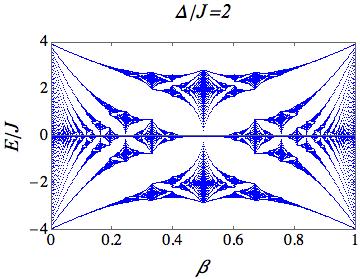}
\caption{Energy spectrum of the Aubry-Andr\'e model as a function of $\beta$ for two different values of $\Delta/J$. At the transition, $\Delta/J=2$, the spectrum gives rise to the Hofstadter butterfly.}
\label{Fig8}
\end{figure}

To conclude this section, in \fref{Fig8} we show the energy spectrum of the Aubry-Andr\'e model as a function of the incommensurate parameter $\beta$ for two different disorder strengths $\Delta/J$. As described above, the values of $\beta$ must be chosen such that they have some degree of incommensurability. Here we have used $\beta \in [0,1)$ such that $\beta= i/\Omega$ with $i =1, 2, 3 ,... , 987$. This spectrum has been studied in numerous works \cite{Sokoloff,Azbel,Hofstadter, Kraus,Madsen,Fangli} since it displays very rich structure in both the extended and localized regimes.

For values of disorder amplitude above and below of the transition point, the energy spectrum structure is completely different from that associated to $\Delta/J=2$, where it shows the Hofstadter butterfly spectrum \cite{Hofstadter}. The case $\Delta/J=2$ has been widely investigated, since it describes the quantum physics of an electron moving on a two dimensional square lattice in the presence of a transverse magnetic field \cite{Harper}.\\

\section{Time dependent results}
In this section, we discuss the time evolution of a given initial condition, in presence of disorder. This is one of the ways in which experimentalists measure how much a system apart from an initially delocalized or extended state \cite{Roati1,Bloch2}. First, we briefly summarize some basic concepts related to the evolution in time of single particle problems. According to quantum mechanics, the time evolution of a ket $|\psi\rangle$ is given by the time dependent Schr\"odinger equation:
\begin{eqnarray}
i\hbar\frac{\partial}{\partial t}|\psi\rangle = H|\psi\rangle.
\label{eq23} 
\end{eqnarray}
Again, we expand the ket in the site basis $|\psi\rangle= \sum_{i} \psi_{i}|x_{i}\rangle$ and after substitution in the equation (\ref{eq23}), we obtain the time dependent equation for the coefficients $\psi_{i}$
\begin{eqnarray}
\frac{d}{d t}\psi_{i} = -\frac{i}{\hbar}\sum_{j}H_{ij}\psi_{j},
\label{eq24} 
\end{eqnarray}
where $H_{ij}$ are the matrix elements of $\mathbf{H}$ given in equality (\ref{eq20}). Equation (\ref{eq24}) represents a system of $\Omega$ coupled ordinary differential equations which can be easily solved by using the Runge-Kutta 4 (RK4) method:
\begin{eqnarray}
\eqalign{\vec{k_{1}} =\mathbf{H}(\tau_{j})\vec{\psi}(\tau_{j})\cr
\vec{k_{2}} =\mathbf{H}(\tau_{j}+\Delta \tau/2)(\vec{\psi}(\tau_{j})+\vec{k}_{1}/2)\cr
\vec{k_{3}} =\mathbf{H}(\tau_{j}+\Delta \tau/2)(\vec{\psi}(\tau_{j})+\vec{k}_{2}/2)\cr  
\vec{k_{4}} =\mathbf{H}(\tau_{j}+\Delta \tau)(\vec{\psi}(\tau_{j})+\vec{k}_{3})\cr    
\psi(\tau_{j+1}) = \psi(\tau_{j}) + \frac{\Delta \tau}{6}(k_{1}+2k_{2}+2k_{3}+k_{4}).
}
\label{eq25} 
\end{eqnarray}
Where we have set $\tau = J t/\hbar$ as our dimensionless unit of time. The advantage of using the RK4 method resides on one side in its accuracy $(\Delta \tau)^{4}$, and on the other side, in the relatively simple way in which the above equations can be implemented. In our calculations, we set $\Delta \tau = 0.01$ which displays conservation of both, the norm and the energy within the whole numerical time evolution.
 
\begin{figure}[h]
\centering
\includegraphics[width=0.4\textwidth]{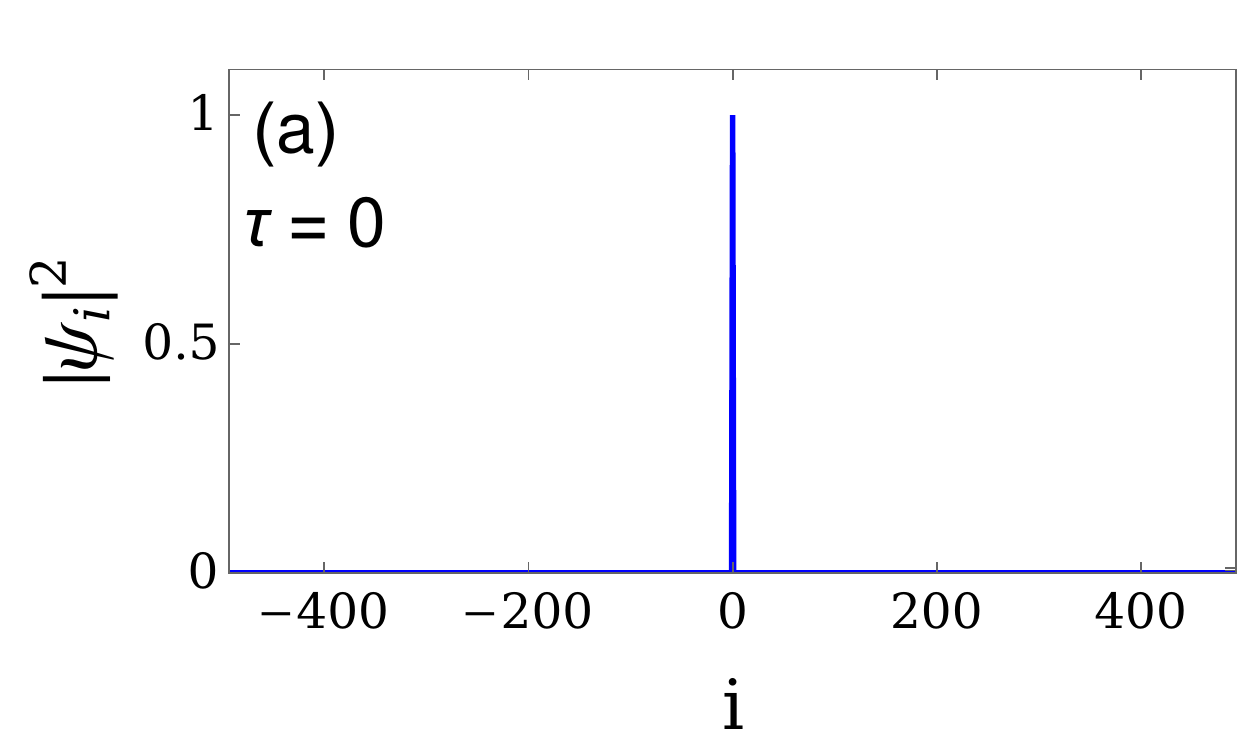}
\includegraphics[width=0.42\textwidth]{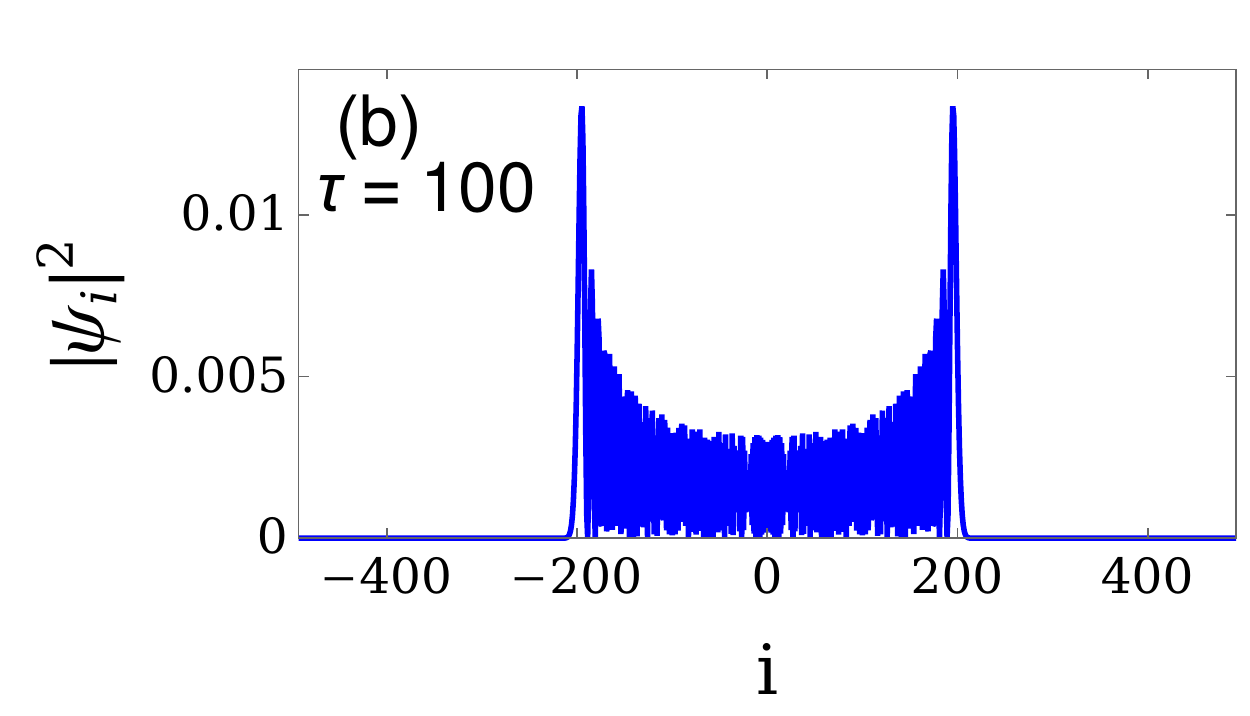}
\caption{(a) Initial density profile localized in the middle of the lattice. (b) Density profile at time $\tau = 100$ in absence of disorder.}
\label{Fig9}
\end{figure}

As it is well known, the time evolution of an eigenstate of $H$ would give trivial results. Nevertheless, the evolution of an arbitrary state can yield signatures of the presence of disorder. For this reason, we first study the  evolution in time of an initially fully localized state in the middle of the lattice $|\psi(\tau=0)\rangle = |x_{0}\rangle$. This initial condition mimics "designs " prepared in current experiments performed with ultracold atoms \cite{Roati1}. In \fref{Fig9}(a) and \fref{Fig9}(b), we plot the initial density profile and the spreading of such initial state in the absence of disorder for a time of $\tau=100$ respectively. 
 
\begin{figure}[h]
\centering
\includegraphics[width=0.7\textwidth]{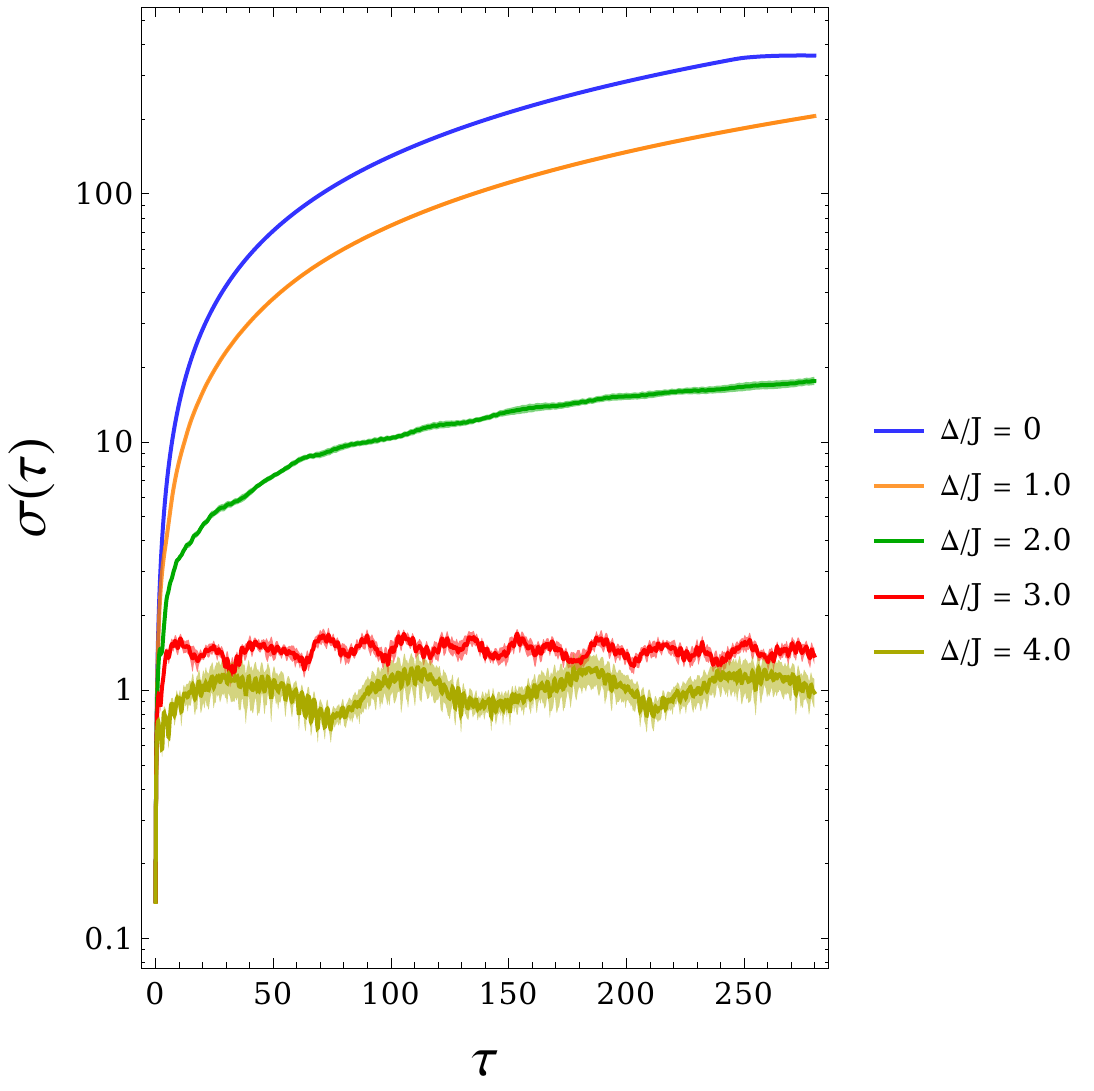}
\caption{Root mean square of the displacement $\sigma$ as a function of time for different disorder strengths $\Delta/J$. The shaded area around in each curve represents the standard deviation over 10 samples of the random phase $\phi$.}
\label{Fig10}
\end{figure}

In order to quantify the spreading of the initial wave function, we determine the root mean square of the displacement (RMSD) in each time step, this latter quantity is defined as 
\begin{eqnarray}
\sigma(\tau) =\left[\sum_{i}i^2|\psi_{i}(\tau)|^2 \right]^{1/2}.
\label{eq26}
\end{eqnarray}
In \fref{Fig10} we plot the RMSD, in logarithmic scale, as a function of the time $\tau$ for five different disorder strengths. The shaded area in each curve represents the standard error over $10$ realizations of random phase $\phi$. As one can observe, for zero disorder $\Delta/J = 0$ the wave packet propagates ballistically, that is  $\sigma \propto \tau$, showing that the RMSD grows linearly in time. Due to the system finite size, the RMSD reach a maximum value and oscillates around it. For this reason, we let the system evolve until the RMSD reaches its maximum value for zero disorder. 

\begin{figure}[h]
\centering
\includegraphics[width=0.55\textwidth]{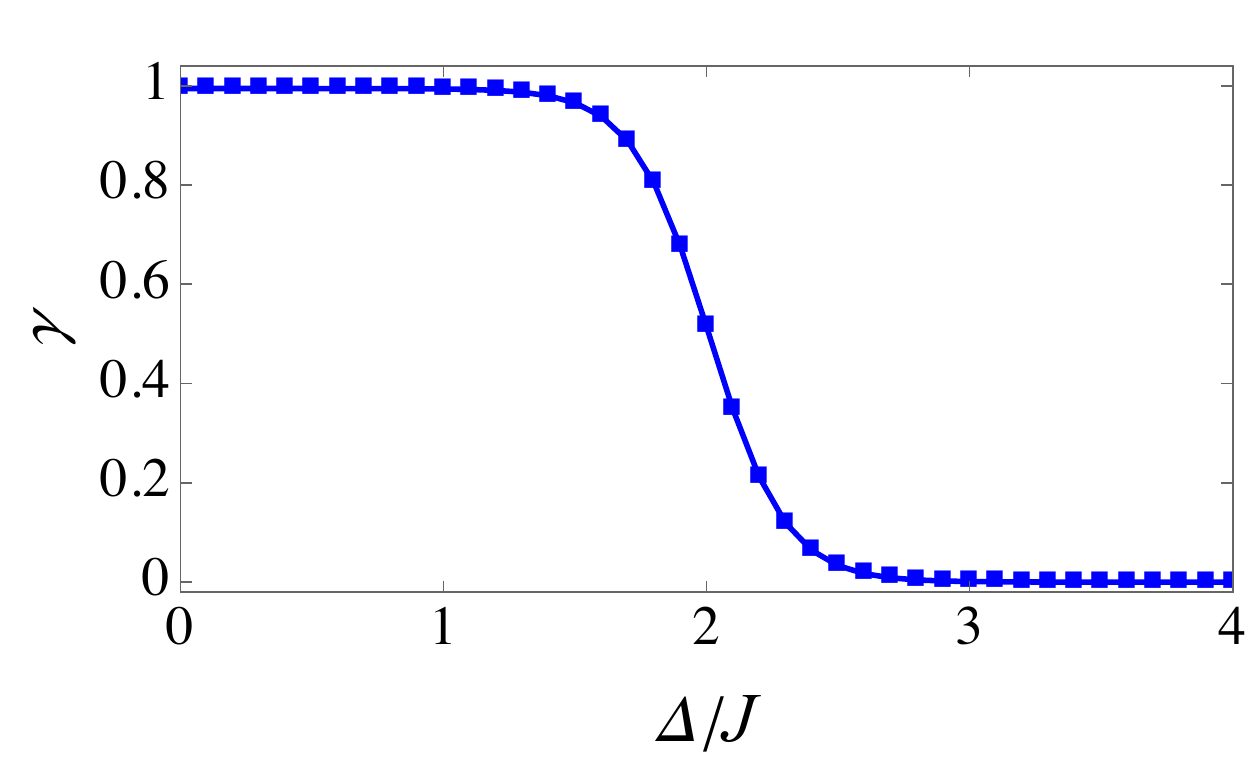}
\caption{Values of the fitted $\gamma$ in the time dependence of the RMSD $\sigma\propto\tau^{\gamma}$ as a function of the disorder strength $\Delta/J$.}
\label{Fig11}
\end{figure}

The time dependence of the RMSD can be fitted with a power law ansatz $\sigma \propto \tau^{\gamma}$. This fit must be done at intermediate times scales, where one neglects the contribution of the transient behavior at short times and the maximum spreading at later ones \cite{Pedro}. \Fref{Fig11} shows the exponent $\gamma$ of the above fit as a function of the disorder strength $\Delta/J$. This plot allows us to identify the ballistic regime $\gamma=1$, the superdiffusive $1/2<\gamma<1$, the subdiffusive $0<\gamma<1/2$,  and the localized one $\gamma = 0$, associated to the diffusion of an initially localized wave packet.\\

\begin{figure}[h]
\centering
\includegraphics[width=0.55\textwidth]{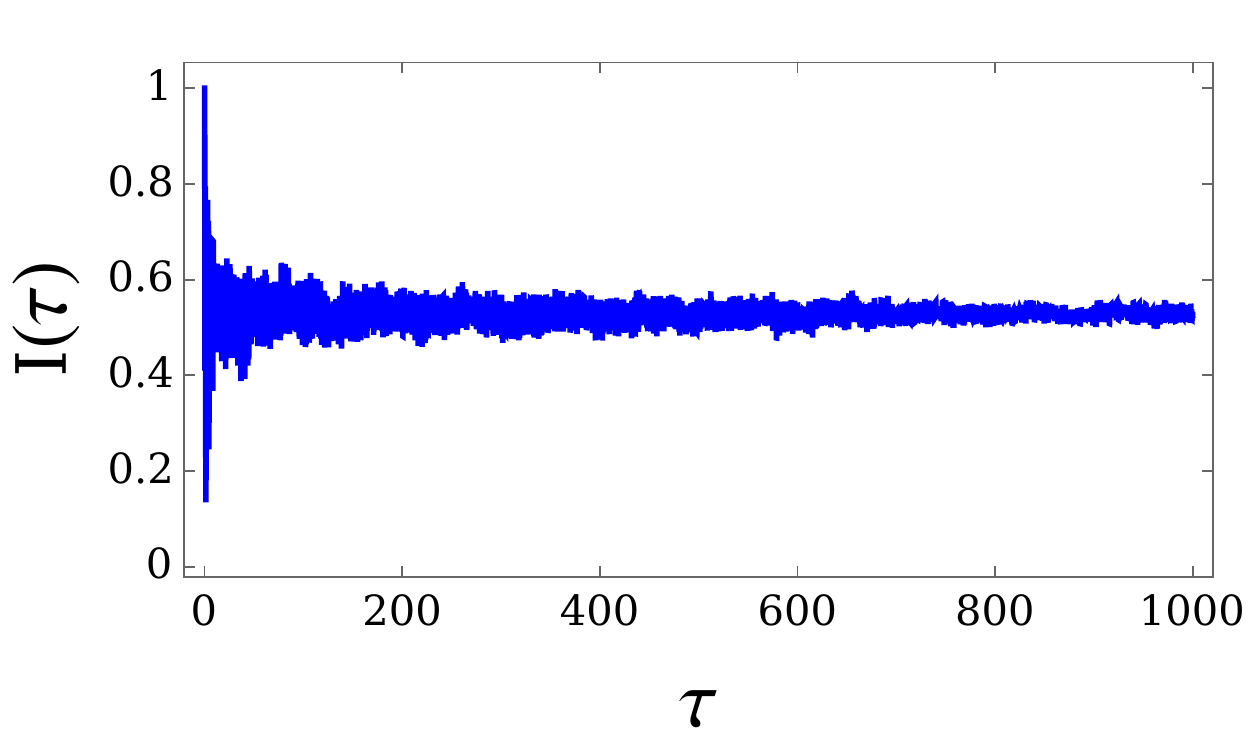}
\caption{Imbalance evolution $I(\tau)$ as a function of time $\tau$ for a disorder strength of $\Delta/J$ = 4.}
\label{Fig12}
\end{figure}

Another interesting observable that allows us to discern between a localized and extended phase,  and can also be detected in current experiments \cite{Bloch3,Bloch2} is the imbalance $I(\tau)$. For a single particle problem, the imbalance parameter is defined as follows:
\begin{eqnarray}
\mathrm{I}(\tau) = \frac{n_{e}(\tau)-n_{o}(\tau)}{n_{e}(\tau)+n_{o}(\tau)},
\label{eq27}
\end{eqnarray} 
where $n_{e}(\tau) = \sum_{i\in even} |\psi_{i}(\tau)|^{2}$ is the total probability density of the particle on even sites and $n_{o}(\tau)=\sum_{i\in odd} |\psi_{i}(\tau)|^{2}$ corresponds to the total probability density of the particle on odd sites of the lattice. In order to condense the whole time evolution of the imbalance, we defined $I$ as the asymptotic value of $I(\tau)$ \cite{Romito}
\begin{eqnarray}
\mathrm{I} = \lim_{\tau\rightarrow \infty} \frac{1}{\tau}\int d\tau '  \ \frac{n_{e}(\tau ')-n_{o}(\tau ')}{n_{e}(\tau ')+n_{o}(\tau ')}
\label{eq27b}
\end{eqnarray} 
For calculation purposes, our numerical simulations were performed until $\tau = 1000$, which corresponds to a real time of $t = 1000 \hbar/J$ that is much bigger than the hopping time in the lattice.\\
Besides being a measurable observable, the most significant advantage of using the imbalance parameter as an order parameter for probing localization is that, it can also provide signatures of many body localization \cite{Jesko}, when interactions are present.
To show the dynamical behavior of $I(\tau)$, we start by considering the evolution of a density-wave like pattern in which only even sites are initially occupied. To have meaningful calculations we have to impose the same number of odd and even sites in the full lattice. Here, we consider $\Omega=1000$ and the incommensurate parameter $\beta= 987/1597$. Then, we calculate the value of the imbalance $I(\tau)$ at every time step until $\tau = 1000$ at which we observe that the imbalance oscillates around its asymptotic value. For illustration purposes in \fref{Fig12}, we plot the imbalance as a function of time $\tau$ for a disorder $\Delta/J = 4.0$

\begin{figure}[h]
\centering
\includegraphics[width=0.55\textwidth]{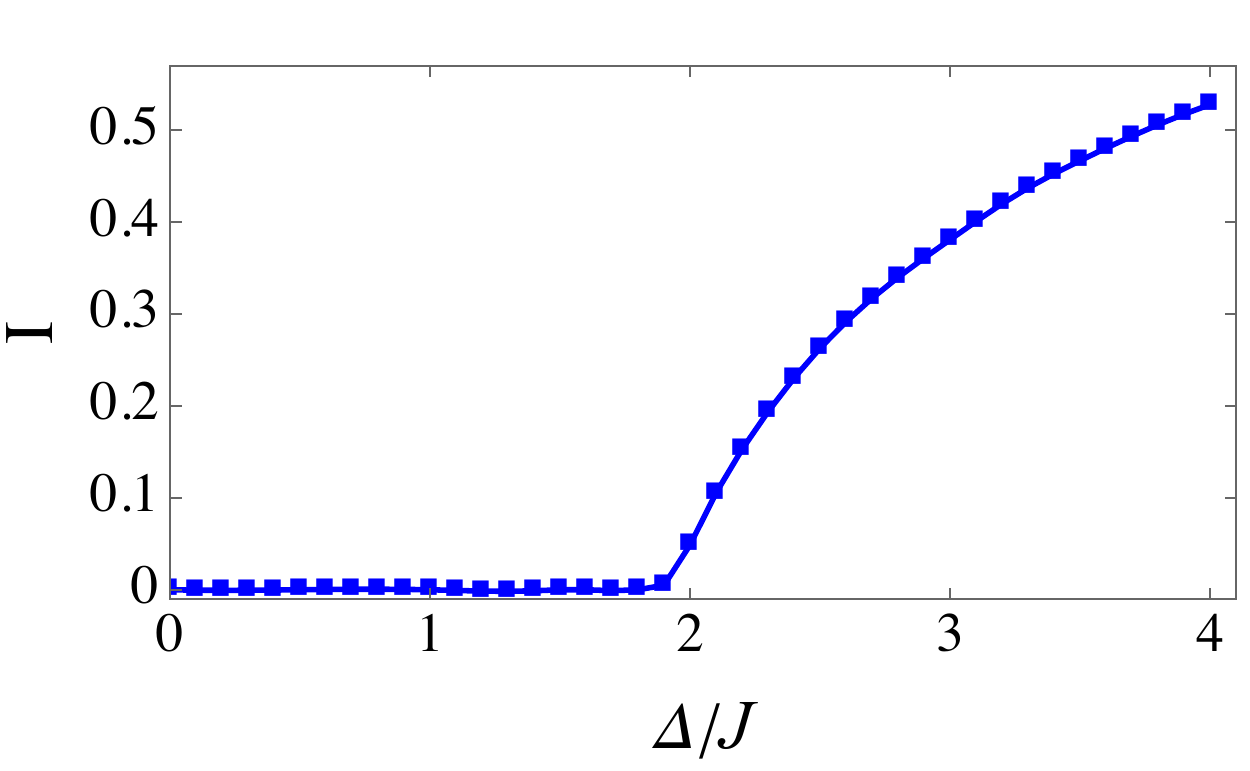}
\caption{The asymptotic value of the imbalance $I$ as a function of the disorder strength $\Delta/J$. The time of evolution was set to $\tau = 1000 \hbar/J$ and the lattice size considered is $\Omega=1000$.}
\label{Fig13}
\end{figure}

In \fref{Fig13} we show the asymptotic value of the imbalance as a function of the disorder strength $\Delta/J$. Since for $\Delta/J<2$ all the states are extended, the particle can easily tunnel to nearby sites leading to a zero value of the imbalance in a few tunneling times.
The vanishing of the imbalance must be interpreted as an indication of ergodicity since the system completely loses any previous information associated with the initial state.
However, for $\Delta/J>2$ the imbalance reaches a finite value which is closer to the initial value as the disorder is increased. This suggests that the system is non-ergodic as it retains certain memory of the initial configuration.  

\section{Final Remarks}

The main purpose of the manuscript is to introduce the undergraduate student to one of the most studied topics in condensed matter, the localization phenomenon. For this purpose, we have presented a comprehensive study of a single particle moving in a disordered lattice in one dimension. In particular, the disorder here analyzed corresponds to a quasiperiodic one. Considering as a starting point a quantum analog of such condensed matter system, namely a weakly interacting ultracold Bose gas confined in a 1D quasiperiodic lattice, a straightforward derivation of the Aubry-Andr\'e Hamiltonian representing this type of disorder was presented. Then, we proceeded to review the essential results of the one-dimensional Aubry-Andr\'e model, as well as the form in which localization transition can be recognized. First, we focused on the description of the Aubry-Andr\'e model properties and then, to characterize the localization transition as a function of the disorder amplitude, we investigated stationary and dynamical properties. Particularly, we determine the inverse participation ratio (IPR) and its opposite, the normalized participation ratio (NPR) that provide information of how localized across the lattice a wave function is, as a function of the disorder amplitude. These parameters were calculated for both, the ground state and the full energy spectrum exhibiting that the spectrum is either, completely delocalized or completely localized, but not a mixture of both, localized and extended states. Another stationary property here studied was the energy spectrum that exhibits the classical Hofstadter pattern when the disorder amplitude reaches a critical value.

All the formalism and techniques used here are at the level of an advanced undergraduate student or equivalent. We believe that this kind of readings brings a student to become closer to the  comprehension of current research on single particle and many particle localization phenomena. With the tools used in this paper, the interested reader can address the study of vanguard problems related to the central topic of this manuscript, for instance, the dependence of localization on dimensionality, the competition of disorder and interactions, the effects of next nearest neighbors in the localization phenomenon, and the response of the system in the novel driven disordered lattices among others. Indeed, the effects of disorder combined with strong interactions remains an outstanding challenge to the theory.

Finally, in this paragraph we want to briefly summarize some of the reported predictions of the Aubry-Andr\'e model. These include spin-orbit coupling effects \cite{Rajesh}, closed expressions for the energy separating localized and non-localized states \cite{Liu-Gao} and coexistence of localized and extended states in interacting quasiperiodic systems \cite{Yucheng} among others. At the many body level, localization of the ground state established rigorously in the weakly interacting regime for both, repulsive and attractive interactions \cite{Mastropietro} and many-body localization vs. thermalization and onset of equilibrium \cite{Vadim}, which can have implications for quantum devices and quantum computation. All of these phenomena can be analyzed with the same tools and properties here presented.

\section*{Acknowledgments}
We acknowledge useful discussion with G. Garc\'ia Naumis. GADC acknowledges CONACYT scholarship. This work was partially funded by grant INI105217 DGAPA (UNAM) and 255573 CONACYT.

\end{document}